\documentclass[manuscript]{aastex}

\slugcomment{accepted by ApJL}

\shorttitle{Physical Properties of Near-Earth Asteroid 2011~MD}
\shortauthors{Mommert et al.}

\begin{document}


\title{Physical Properties of Near-Earth Asteroid 2011~MD}


\author{M. Mommert}
\affil{Department of Physics and Astronomy, Northern Arizona
  University, PO Box 6010, Flagstaff, AZ 86011, USA}

\author{D. Farnocchia}
\affil{Jet Propulsion Laboratory, California Institute of Technology,
  Pasadena, CA 91109, USA}

\author{J.~L. Hora}
\affil{Harvard-Smithsonian Center for Astrophysics, 60 Garden Street,
  MS 65, Cambridge, MA 02138-1516, USA}

\author{S.~R. Chesley} \affil{Jet Propulsion Laboratory, California
  Institute of Technology, Pasadena, CA 91109, USA}


\author{D.~E. Trilling}
\affil{Department of Physics and Astronomy, Northern Arizona
  University, PO Box 6010, Flagstaff, AZ 86011, USA}

\author{P.~W. Chodas} \affil{Jet Propulsion Laboratory, California
  Institute of Technology, Pasadena, CA 91109, USA}

\author{M. Mueller} 
\affil{SRON Netherlands Institute for Space
  Research, Postbus 800, 9700 AV, Groningen, The Netherlands}

\author{A.~W. Harris}
\affil{DLR Institute of Planetary Research, Rutherfordstrasse 2, 12489
  Berlin, Germany}

\author{H.~A. Smith}
\affil{Harvard-Smithsonian Center for Astrophysics, 60 Garden Street,
  MS 65, Cambridge, MA 02138-1516, USA}

\and

\author{G.~G. Fazio}
\affil{Harvard-Smithsonian Center for Astrophysics, 60 Garden Street,
  MS 65, Cambridge, MA 02138-1516, USA}



\begin{abstract}
  We report on observations of near-Earth asteroid 2011~MD with the
  {\it Spitzer Space Telescope}. We have spent 19.9~h of
  observing time with channel 2 (4.5~$\mu$m) of the Infrared Array
  Camera and detected the target within the 2$\sigma$ positional
    uncertainty ellipse. Using an asteroid thermophysical model and a
  model of nongravitational forces acting upon the object we constrain
  the physical properties of 2011~MD, based on the measured flux
  density and available astrometry data. We estimate 2011~MD to be
  (6$^{+4}_{-2}$)~m in diameter with a geometric albedo of
  0.3$^{+0.4}_{-0.2}$ (uncertainties are 1$\sigma$). We find the
  asteroid's most probable bulk density to be
  (1.1$^{+0.7}_{-0.5}$)~g~cm$^{-3}$, which implies a total mass of
  (50--350)~t and a macroporosity of ${\geq}$65\%, assuming a material
  bulk density typical of non-primitive meteorite materials. A
    high degree of macroporosity suggests 2011~MD to be a rubble-pile
  asteroid, the rotation of which is more likely to be retrograde than
  prograde.
\end{abstract}


\keywords{minor planets, asteroids: individual (2011 MD) --- infrared: planetary systems}



\section{Introduction}

Little is known about the physical properties of near-Earth asteroids
with diameters smaller than 100~m. \citet{Mainzer2014} measured the
sizes and albedos of the smallest optically discovered near-Earth
asteroids ($d > 10$~m) from NEOWISE data. \citet{Mommert2014}
constrained a number of physical properties of candidate mission
target 2009~BD, revealing two extraordinary but equally possible
solutions. 

Near-Earth asteroid 2011~MD was discovered on June 22, 2011, by the
Lincoln Near Earth Asteroid Research (LINEAR) program
\citep{Blythe2011}. Five days later, the object passed Earth within a
distance of 15000~km from the surface, which significantly
changed the object's orbit. 2011~MD now has a specific linear momentum
($\Delta v$), the launch velocity necessary to reach 2011~MD with
spacecraft, of 4.17~km~s$^{-1}$ (Larry H. Wasserman, personal
communication 2014), making it a very accessible candidate space-mission
target asteroid. Photometric time series revealed a rotational period
of (0.1939$\pm$0.0004)~h with a peak-to-peak amplitude of 0.8~mag
\citep{Ryan2012, CALL}. Based on its absolute magnitude
\citep[$H=28.0\pm0.3$,][]{MPC,Horizons,Micheli2014}, the apparent
magnitude of an asteroid at 1~au from the Sun and the observer at a
solar phase angle of zero, and assuming a most probable albedo range
of 0.03--0.50, its possible diameter ranges between 4 and 22 m.


2011~MD is a potential candidate for NASA's proposed Asteroid Robotic
Redirect Mission \citep[ARRM,][]{NASAAsteroidInitiativeWebsite,
  Mazanek2013}. One of the proposed mission concepts for ARRM involves
capturing an asteroid less than ${\sim}$10~m in size and guiding it
into orbit about the Moon, where it could be visited and explored by
astronauts. Candidate asteroids for this concept could have masses in
the range of tens to hundreds of metric tons, but the maximum mass for
each candidate would depend on its orbital parameters. The size and
mass of 2011~MD were not known accurately enough to say whether it
could be considered a more serious candidate for the proposed mission.

We utilize observations obtained by the {\it Spitzer
  Space Telescope} to constrain the physical properties of 2011~MD.

\section{{\it Spitzer} Observations and Data Reduction}
\label{lbl:observations}

\begin{figure}
\epsscale{.80}
\plotone{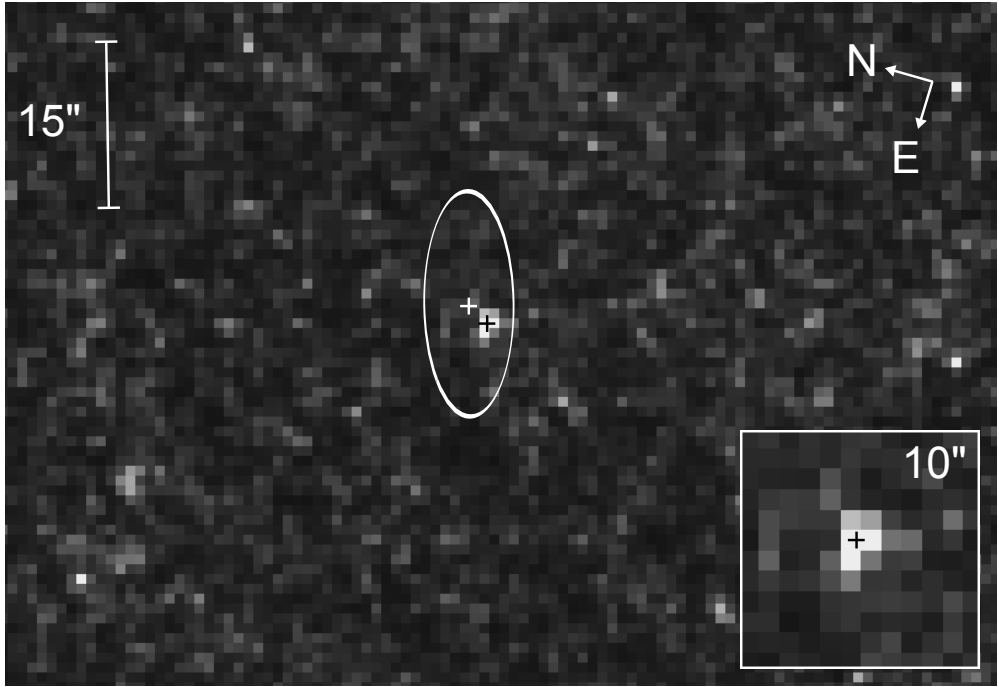}
\caption{IRAC channel 2 (4.5~$\mu$m) map centered and stacked in the
  co-moving frame of 2011~MD, using a power-law color table
    scaling. The object's position (black cross) lies within
    1$\sigma$ in RA and 2$\sigma$ in Dec of the predicted position
    (white cross). The white ellipse depicts the 3$\sigma$ positional
    uncertainty. We derive a flux density and 1$\sigma$ uncertainty
  of ($0.60\pm0.27$)~$\mu$Jy. The inset shows a
  10\arcsec$\times$10\arcsec\ postage stamp of
  2011~MD. \label{fig:map}}
\end{figure}

We observed 2011~MD with the Infrared Array Camera
\citep[IRAC,][]{Fazio2004} on-board the {\it Spitzer ​Space Telescope}
\citep{Werner2004} in Program ID 10132 using a total of 19.9~h
of observation time. Observations (astronomical observation request
49716480) started on 11 February, 2014, 20:30:47~UT, using the
``Moving Single'' object mode to track in the moving frame of the
object.  We performed the observations in full array mode with 100~s
frames in IRAC channel 2 (4.5~$\mu$m) only, using a medium
cycling dither pattern with 227 dither positions and 3 repeats, 
  resulting in a  total of 681 frames, or 18.3~h on-source
exposure time.

At the time of the observations, 2011~MD was 1.09~au from the
Sun and 0.14~au from {\it Spitzer} with a solar phase angle of
54$^\circ$. The observation window was selected based on {\it Spitzer}
observing constraints.

The data were reduced using the method by \citet{Mommert2014}. A
mosaic of the field is​ constructed from the dataset itself and then
subtracted from the individual basic calibrated data (BCD)
frames. After subtraction of the background mosaic, residual
background sources and bright cosmic ray artifacts are masked in the
individual BCDs before being mosaicked in the reference frame of the
moving object.

In the final co-move map we find a source within 2$\sigma$ of the
  expected position of 2011~MD (see Figure \ref{fig:map}, and Sections
  \ref{lbl:discussion} and \ref{lbl:modeling} for a discussion). We
identify this source as 2011~MD and derive a flux density of
(0.60$\pm$0.27)~$\mu$Jy. The uncertainty is derived as the standard
deviation of the photometry of implanted fake sources with flux
densities of 0.6~mJy into various positions of the co-move map.

\section{Modeling}
\label{lbl:modeling}

We constrain the physical properties of 2011~MD by combining an
asteroid thermophysical model with a model of the nongravitational
forces acting on the asteroid, similar to the approach taken by
\citet{Mommert2014}.

The thermophysical model approximates the surface temperature
distribution of 2011~MD and is used to determine the thermal-infrared
emission from its surface as a function of its physical
  properties, including spin axis orientation (represented by the
obliquity, $\gamma$), rotational period, $P$, thermal inertia,
$\Gamma$, and surface roughness. Surface roughness causes infrared
beaming, an effect that focuses thermal emission radiated towards the
observer, and is modeled as emission from spherical craters
\citep[see][for more details]{Mueller2007}. The model solves the heat
transfer equation numerically for a large number of plane surface
facets that form a sphere. The model we use is nearly identical to the
one used by \citet{Mueller2007}. Since the single-band nature of
  our observation precludes a direct fit of the target's spectral
  energy distribution, we take a probabilistic approach in which we
  explore the parameter space by varying the individual input
  parameters. 

Similar to \citet{Mommert2014}, we model the nongravitational
acceleration of the object as a result of the solar radiation pressure
\citep[using the approach by][]{Vokrouhlicky2000} and the Yarkovsky
force \citep{Vokrouhlicky2000b}. The model asteroid is assumed to be
spherical and the heat transfer is solved analytically using the
linearized heat transfer equation \citep{Vokrouhlicky1998,
  Vokrouhlicky1999}. By fitting all available astrometric data of
2011~MD, the model derives the bulk density, $\rho$, and the
goodness-of-fit parameter $\chi^2$ as a function of the 
  asteroid's properties.

Ground-based astrometric observations of 2011~MD cover the date range
2011-06-21 to 2011-09-03 (1555 observations) in addition to our
Spitzer detection (2014-02-11). The majority of the observations
were collected during the close Earth encounter of June 2011. Of
  special importance for the deductions made in this work are the
  astrometric measurements performed by \citet{Micheli2014}, which
  extend the observed arc until September 2011. We model the
nongravitational perturbations as
\begin{equation}
\mathbf a_{NG} = (A_1 \hat{\mathbf r} + A_2 \hat{\mathbf t})
\left(\frac{1 \ \mathrm{au}}{r}\right)^2\ ,
\end{equation}
where $\hat{\mathbf r}$ and $\hat{\mathbf t}$ are the radial and
transverse directions, respectively, and $r$ is the heliocentric
distance. $A_2/r^2$ translates into the transverse component of the
Yarkovsky effect \citep{Bottke2006}, whereas $A_1/r^2$ models the
solar radiation pressure and the radial component of the Yarkovsky
effect. We use this simplified model approach for ephemeris
predictions and to investigate the detectability of nongravitational
forces in the astrometric data. 
In order to fit the model to the astrometric data, we applied the
\citet{Chesley2010} debiasing and weighting scheme. Since timing
errors are more relevant when an object is observed at small
geocentric distances, we relaxed the data weights for these
observations. In the case of our Spitzer observations, we applied an
uncertainty of 1\arcsec, accounting for the positional uncertainty of
2011~MD from our observation and the uncertainty of Spitzer's
ephemeris (J.\ Lee and T.~J.\ Martin-Mur, personal communication
2014). The orbital fit (JPL Solution 40) to the observations yields
$A_1 = (7.21 \pm 2.26) \times 10^{-11}$ au/d$^2$ (3.4$\sigma$
confidence) and $A_2 = (-1.13 \pm 2.91) \times 10^{-12}$ au/d$^2$
(0.4$\sigma$). Our value of $A_1$ agrees within uncertainties with the
value found by \citet{Micheli2014} ($A_1 = (7.3\pm1.4)\times
10^{-11}$ au/d$^2$, 5.2$\sigma$). We ascribe our higher uncertainty to a less
strict weighting used for some of the available astrometric data, and
the fact that we have taken into account the Yarkovsky effect ($A_2$),
which was neglected by \citet{Micheli2014}, and leads to additional
uncertainty, due to the correlation of $A_1$ and $A_2$.

\section{Results}
\label{lbl:results}

We explore the physical property space of 2011~MD based on our flux
density measurement, using a Monte Carlo method in which we generate
40000 randomized synthetic objects. We sample the rotation period
$P=(0.1939\pm0.0004)$~h \citep{Ryan2012, CALL}, the absolute
magnitude $H=28.0\pm0.3$ \citep{Micheli2014}, and the photometric
slope parameter $G=0.18\pm0.13$ (average from all $G$
measurements of asteroids, see \citet{SBDSE}), using normal
distributions. Due to the lack of observational constraints, we
uniformly sample the physically meaningful ranges in
thermal inertia (10--5000~SI units, where 1~SI unit equals
  1~J~m$^{-2}$~s$^{-0.5}$~K$^{-1}$), the azimuth of the spin axis
orientation and the cosine of the obliquity (covering $\gamma$ =
0--180$^\circ$, sampling the cosine leads to a truly random
distribution of the spin vector), and use various surface roughness
models \citep[see,][]{Mueller2007}. We draw flux densities from a normal distribution
with a mean of 0.6~$\mu$Jy and a 1$\sigma$ uncertainty of 0.27~$\mu$Jy
(we reject negative flux densities). For each set of input parameters,
the diameter and albedo are derived by fitting the thermophysical
model flux density to a randomized flux density. The resulting
diameters and albedos, as well as the input parameters
of the thermophysical model are then used in the orbital model in
order to derive $\rho$ and $\chi^2$ for each synthetic object. The
final distributions in diameter, albedo, obliquity, and density are
weighted using $\chi^2$ from the orbital fit in order to account for
the compatibility with the astrometric data. Other parameters, like
$H$, $G$, and $\Gamma$, are not sensitive to $\chi^2$.

We reject synthetic model asteroids with unphysically high Bond
albedos. The Bond albedo, $A$, describes the reflectivity integrated
over the whole electromagnetic spectrum and can be approximated as $A
\sim q \cdot p_v$, with the phase integral $q = 0.290 + 0.684 \cdot G$
\citep{Bowell1989}. Hence, high geometric albedos $p_V$ can lead
to $A>1$, which contradicts the law of conservation of energy.

\begin{figure}
\epsscale{.80}
\plotone{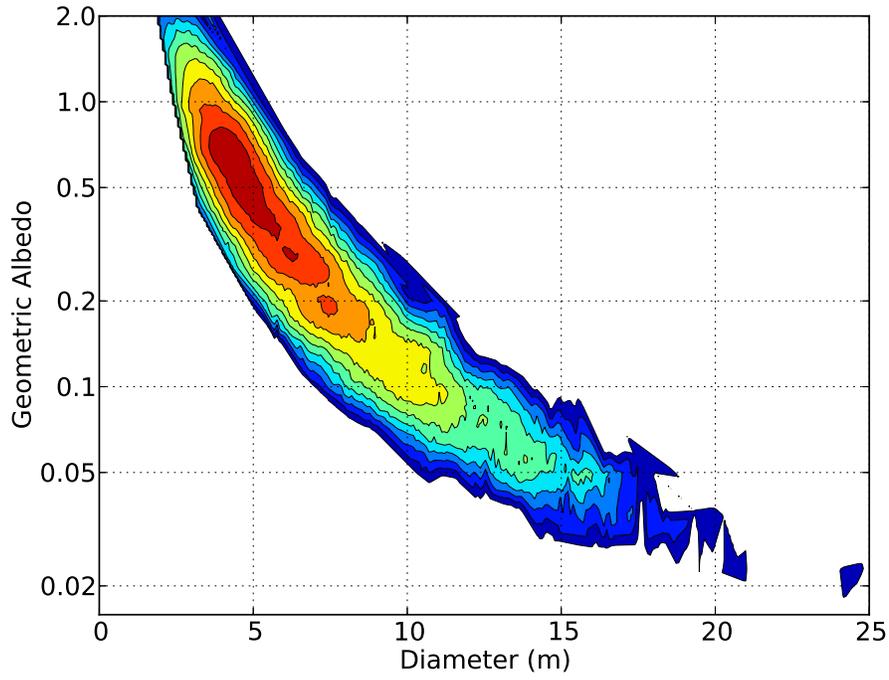}
\caption{Distribution of the 40000 synthetic model asteroids generated
  in the Monte Carlo method in albedo-diameter space. Contour lines
  and colors represent the logarithm of the weighted number density of
  synthetic model asteroids per space element. The median of the
  distributions in diameter and albedo is 6~m and 0.3,
  respectively.\label{fig:diameter_albedo}}
\end{figure}

\begin{figure}
\epsscale{.80}
\plotone{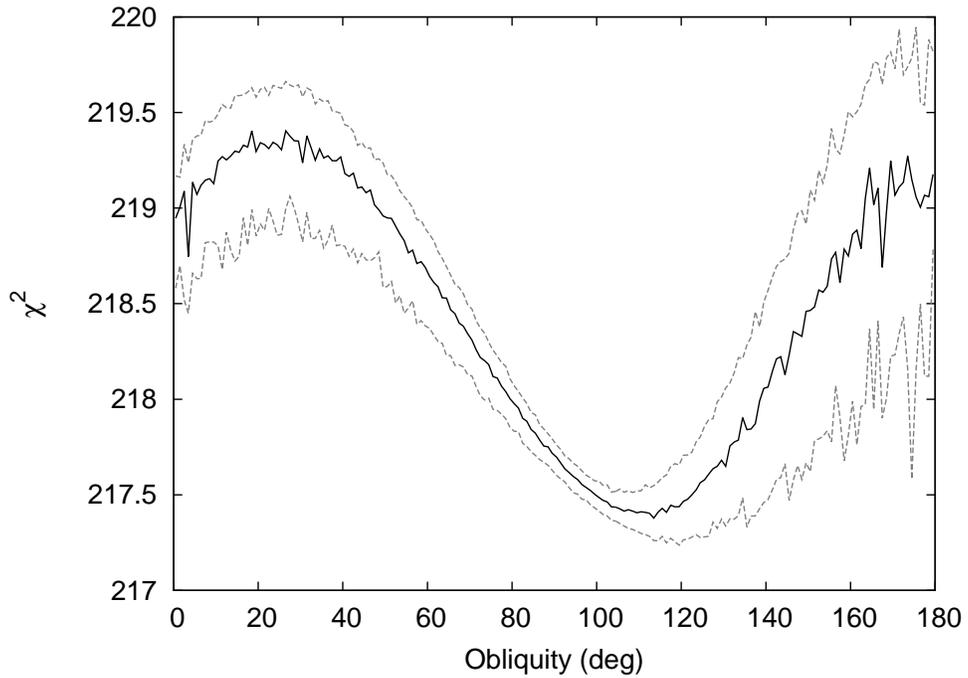}
\caption{Orbital model $\chi^2$ distribution as a function of
  obliquity as derived from the Monte Carlo method. The continuous
  black line represents the median per bin in obliquity; dashed gray
  lines indicate the 1$\sigma$ confidence interval. The distribution
  shows a global minimum at ${\sim}$110\degr, favoring a retrograde
  rotation of 2011~MD.\label{fig:obl_chi2}}
\end{figure}

\begin{figure}
\epsscale{.80}
\plotone{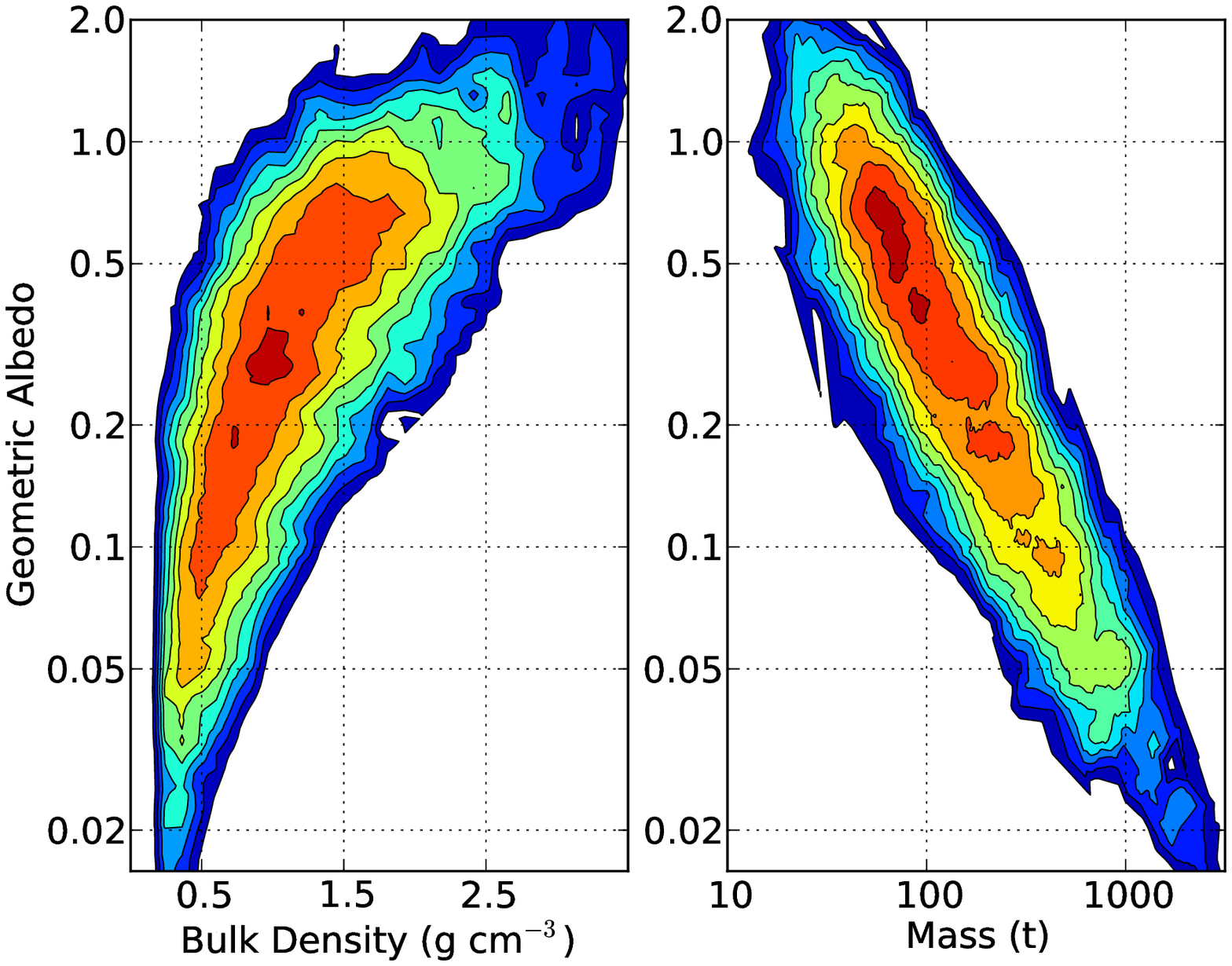}
\caption{Bulk density (left) and total mass (right) distributions of
  the synthetic model asteroids, weighted with the orbital model
  $\chi^2$ (see Figure \ref{fig:diameter_albedo} for
  definitions). Both parameters are a strongly correlated to
  albedo. We adopt the median of each weighted distribution, yielding
  a most probable bulk density of (1.1$^{+0.7}_{-0.5}$)~g~cm$^{-3}$ and
  total mass of ($110^{+240}_{-60}$)~t.\label{fig:density_mass}}
\end{figure}

The final results of our analysis are depicted in Figures
\ref{fig:diameter_albedo} to \ref{fig:density_mass} and show that
2011~MD has a mean diameter of (6$^{+4}_{-2}$)~m (1$\sigma$) and an
albedo of $0.3^{+0.4}_{-0.2}$ (1$\sigma$). Note that in the case
  of asymmetric uncertainties, the 1$\sigma$ confidence interval
  refers to the 68.3\% of values higher/lower than the median value.
The 3$\sigma$ confidence interval covers a range of (2--26)~m in
diameter and ${\geq}0.02$ in albedo. From the orbital model we find a
most probable bulk density of (1.1$^{+0.7}_{-0.5}$)~g~cm$^{-3}$
(1$\sigma$, 3$\sigma$ interval: (0.2--5.0)~g~cm$^{-3}$), which
translates into a total mass of ($110^{+240}_{-60}$)~t
($1\sigma$, 3$\sigma$ interval: (10--2500)~t). The measured
albedo is compatible with a number of non-primitive taxonomic
  classes \citep{Thomas2011}.

Our model results favor a retrograde rotation of 2011~MD, which is
suggested by the $\chi^2$ distribution produced by the orbital model
(see Figure \ref{fig:obl_chi2}), or the negative value of $A_2$
\citep[compare to][]{Farnocchia2013}. Note that in case of a complex rotation
of the object, our definition of obliquity is referenced to the
rotational angular momentum vector rather than the spin axis. We are
unable to constrain the thermal inertia of 2011~MD, given the low
confidence in the measurement of $A_2$.

\section{Discussion}
\label{lbl:discussion}

Figure \ref{fig:map} depicts the 3$\sigma$ positional uncertainty of
2011~MD during our observations as an ellipse with semimajor axes
9.9\arcsec\ and 3.9\arcsec\ at an angle of 163\degr\ (East to
North). The uncertainty is based on all available ground-based
astrometric data, physically reasonable values of $A_1$ and $A_2$, as
well as Spitzer ephemeris and astrometric image calibration
uncertainties. The position of the source associated with 2011~MD
agrees within less than 2$\sigma$ with the expected postion of 2011~MD
and has the highest signal-to-noise ratio in the co-move map. The
appearance of the source agrees with the IRAC point-spread function,
which has a full width half maximum of 1.66\arcsec\ (1.4~px). Figure
\ref{fig:map} shows that potential sources other than 2011~MD are
constrained to only one high-signal pixel. We investigated the
possibility that the source we identify as 2011~MD is a product of
noise. For this reason, we created co-move maps from the BCDs with
rates that are close, but not identical, to the rate of 2011~MD. This
approach precludes an alignment of the positions of 2011~MD in the
co-move maps, which have noise properties on both large and small
scales that are nearly identical to those of the original co-move
map. We identified potential sources within 1\arcmin\ of the image
center in each map and compared their flux densities with the
measurements from the original co-move map showing 2011~MD. In a total
of 1000 co-move maps, we found the brightest source to have a signal
that is 20\% lower than that measured for 2011~MD; only 0.01\% of all
potential sources have flux densities that are 20--30\% lower than the
flux density measured for 2011~MD. The probability that the source we
identify as 2011~MD is a noise feature and falls within the 3$\sigma$
error ellipse is ${\leq}5\cdot 10^{-6}$.  Further note that
individual BCDs were aligned in the moving frame of 2011~MD. During
the observations, 2011~MD covered a distance of ${\sim}16.6$\arcmin\
(${>}3$ IRAC fields of view), basically ruling out the possibility
that the source is a background object or a moving object on an orbit
different than that of 2011~MD. Hence, we are confident that we
correctly identified 2011~MD.

  The model approach used in this work is identical to the one used by
  \citet{Mommert2014}. Both the thermophysical and the orbital model
  have been tested extensively and compared to other models. We take
  this Monte Carlo approach in order to minimize the number of {\it a
    priori} assumptions on the properties of 2011~MD; e.g., we do not
  preclude high albedos. We allow for albedo up to values where the
  Bond albedo reaches unity (see Section
  \ref{lbl:results}). Restricting the upper-limit further to values
  that have been observed in other asteroids \citep[$p_V < 1.0$, see,
  e.g.,][]{Thomas2011, Mainzer2011} changes our model results only
  slightly and we find a most probable diameter of 6.2~m and a bulk
  density of 1.0~g~cm$^{-3}$. Note that these values are well within
  the 1$\sigma$ confidence intervals of our nominal model solutions.

The wide range of possible albedos precludes a rough taxonomic
classification of 2011~MD. However, it is very unlikely that 2011~MD
is a primitive asteroid type with an albedo less than 0.1; the
probability for $p_V \leq 0.1$ is only 5\%. Assuming 2011~MD to
consist of material comparable to ordinary chondrites, which has the
lowest density of all non-primitive materials, we can derive a lower
limit on the macroporosity of this object of 65\%
\citep[see,][]{Mommert2014, Britt2002}. This high degree of
macroporosity suggests a rubble-pile nature for 2011~MD, which is
possible despite its fast rotation \citep{Scheeres2010}.

Our bulk density estimate (1.1~g~cm$^{-3}$) is nearly twice as high as
the value found by \citet{Micheli2014}. This difference is caused by
their neglect of Yarkovsky forces and the assumption that 2011~MD's
albedo follows the albedo distribution for small ($10 < d < 100$~m)
asteroids \citep{Mainzer2014}. Using the same assumptions, our results
are consistent (M.\ Micheli, personal communication 2014). Figure
\ref{fig:density_mass} plots bulk density and total mass as a function
of albedo, both of which show a strong dependence on albedo.

We compare the physical properties of 2011~MD with those found for
other small asteroids. In comparison to the extraordinary solutions
for 2009~BD \citep{Mommert2014}, 2011~MD is slightly larger (diameter
of 2009~BD: 2.9~m or 4.0~m), has a lower density (higher
macroporosity) than either solution ($\rho$ = 2.9~g~cm$^{-3}$ or
$\rho$ = 1.7~g~cm$^{-3}$), and has a more moderate albedo ($p_V=0.85$
or $p_V=0.45$). The albedo measured for 2011~MD is compatible with the
albedo distribution of 10 to 100~m-sized asteroids found by
\citet{Mainzer2014}. The bulk density and macroporosity of 2011~MD are
comparable to values observed in some asteroids larger than 100~m
\citep[see][]{Mommert2014}.

Note that the diameter derived as part of this work is the effective
diameter of a sphere with the same volume as the real shape of
2011~MD. The large lightcurve amplitude of 0.8~mag \citep{Ryan2012},
however, suggests a highly elongated shape of 2011~MD with an axis
ratio of $b/a\sim$0.5, where ($a$,$b$,$c$) are the axes of a triaxial
ellipsoid. The rotational period of 0.1939~h \citep{Ryan2012} is
significantly shorter than our observation duration (19.9~h);
any optical lightcurve effects are hence averaged over our
observation. We investigate the impact of the temperature distribution
of an ellipsoid with an axis ratio similar to that of 2011~MD compared
to that of a sphere. We use a simplistic model of the shape of 2011~MD
that is based on a triaxial ellipsoid with axis ratio (1,0.5,0.5);
since there is no information on the $c$ axis, we assume $c=0.5$,
which provides a principal axis rotation of the body. We approximate
the measured lightcurve of 2011~MD with a step function: the observer
is faced the long side of the asteroid ($a \times c$) for 75\% of the
rotation period and the short side ($b \times c$) for the remaining
25\% \citep[compare with Figure 2 by][]{Ryan2012}. We realize this
lightcurve behavior by using a composite flux density that consists to
75\% of the flux density emitted by the long side and to 25\% of the
flux density emitted by the short side. We compare the diameter
derived with this composite flux density with that of a spherical
shape and find differences up to 20\%, depending on the spin axis
orientation and thermal inertia. This uncertainty, which is based on a
coarse approximation of the real shape of 2011~MD, is well within the
nominal 1$\sigma$ diameter uncertainties used in our model
approach. Also, the assumed ellipsoidal shape has a cross section that
is different from that of a spherical shape, which affects the solar
radiation pressure acting on the object, and hence changes its bulk
density. We find that the average cross section of the ellipsoid is
10\% larger than that of a sphere, forcing the same change in bulk
density. Again, the nominal uncertainty in bulk density is
significantly larger than this change.

Our observations have provided a determination of the physical
properties of 2011~MD, and in particular, its size and mass. A final
evaluation of 2011~MD as a candidate target for the proposed ARRM
mission is beyond the scope of this work.


\acknowledgments

Some of the computational analyses were run on Northern Arizona
University's monsoon computing cluster, funded by Arizona's Technology
and Research Initiative Fund. M.\ Mommert would like to thank P.\
Penteado for support on the computational aspects of this work. We
would like to thank J.\ Lee and T.~J.\ Martin-Mur for providing
information on the Spitzer ephemeris uncertainties. We thank an
anonymous referee for useful suggestions that led to the improvement
of this manuscript. The work of D.\ Farnocchia, S.\ Chesley, and P.\
W. Chodas was conducted at the Jet Propulsion Laboratory, California
Institute of Technology under a contract with the National Aeronautics
and Space Administration. J.~L.\ Hora and H.~A.\ Smith acknowledge
partial support from Jet Propulsion Laboratory RSA \#1367413. This
work is based on observations made with the {\it Spitzer Space
  Telescope}, which is operated by the Jet Propulsion Laboratory,
California Institute of Technology under a contract with NASA.



{\it Facilities:} \facility{{\it Spitzer}}

\clearpage




\clearpage

\end{document}